\documentclass[conference]{IEEEtran}
\IEEEoverridecommandlockouts
\usepackage{cite}
\usepackage{amsmath,amssymb,amsfonts}
\usepackage{algorithmic}
\usepackage{graphicx}
\usepackage{textcomp}
\usepackage{xcolor}
\usepackage{subfigure}
\usepackage{float}
\usepackage[top=0.76in, bottom=1.01in, left=0.635in, right=0.635in]{geometry}
\def\BibTeX{{\rm B\kern-.05em{\sc i\kern-.025em b}\kern-.08em
    T\kern-.1667em\lower.7ex\hbox{E}\kern-.125emX}}
\begin{document}

\title{Signal Design for AF Relay Systems using Superposition Coding and Finite-Alphabet Inputs\\
	{\thanks{This work of Zheng Dong was supported by the Fundamental Research Funds of Shandong University (61170079614095).}
}
}

\author{\IEEEauthorblockN{Bohai Li\IEEEauthorrefmark{1}, He Chen\IEEEauthorrefmark{2}, Zheng Dong\IEEEauthorrefmark{3}, Yonghui Li\IEEEauthorrefmark{1}, Branka Vucetic\IEEEauthorrefmark{1}}%
		\IEEEauthorblockA{\IEEEauthorrefmark{1}School of Electrical and Information Engineering, The University of Sydney, Sydney, Australia}
		\IEEEauthorblockA{\IEEEauthorrefmark{2}Department of Information Engineering, The Chinese University of Hong Kong, Hong Kong SAR, China}
		\IEEEauthorblockA{\IEEEauthorrefmark{3}School of Information Science and Engineering, Shandong University, China\\
			\IEEEauthorrefmark{1}\{bohai.li, yonghui.li, branka.vucetic\}@sydney.edu.au,
			\IEEEauthorrefmark{2}he.chen@ie.cuhk.edu.hk,
			\IEEEauthorrefmark{3}zhengdong@sdu.edu.cn}
}

\maketitle

\begin{abstract}
This paper focuses on the signal design in a Gaussian amplify-and-forward (AF) relay system with superposition coding (SC) being applied at the relay, which allows the source and relay to transmit their own information within two time slots. Practical quadrature amplitude modulation (QAM) constellations are adopted at the source and relay. To improve the system error performance, we optimize the \textit{weight coefficients} adopted at the source and relay to maximize the minimum Euclidean distance of the received composite constellation, subject to their individual average power constraints. The formulated optimization problem is shown to be a \textit{mixed continuous-discrete} one that is non-trivial to resolve in general. By resorting to the \textit{punched Farey sequence}, we manage to obtain the optimal solution to the formulated problem by first partitioning the entire feasible region into a finite number of sub-intervals and then taking the maximum over all the possible sub-intervals. Simulation results are provided to demonstrate the superior performance of our proposed design based on SC over that using conventional time division multiple access (TDMA).
\end{abstract}

\section{Introduction}
 
Cooperative communication systems have been a long-lasting research topic during the past decades, thanks to its inherent advantages in boosting the achievable rate and communication reliability. Most of the existing cooperative techniques are based on amplify-and-forward (AF) relaying schemes, which make the relay simple, or decode-and-forward (DF) relaying schemes, which avoid noise amplification \cite{b1}. On the other hand, superposition coding (SC) has been extensively studied in the area of information theory to approach the channel capacity \cite{b2}. The embodiment of SC's corresponding information-theoretic ideas in cooperative wireless systems was first proposed in \cite{b3}, where the relay simultaneously transmits its own signal and the signal for which it acts as relay by superimposing them together. We note that most of existing designs for cooperative relay systems using SC are based on Gaussian input signals, whose actual implementation will require unaffordable encoding and decoding efforts \cite{b4}. Furthermore, the actual transmitted signals in practical wireless systems are drawn from finite-alphabet constellations, such as phase shift keying (PSK) modulation and quadrature amplitude modulation (QAM). Applying the results obtained from Gaussian input signals in practical systems using finite-alphabet signals may cause significant performance loss \cite{b5}. 

To the best of our knowledge, there is only a handful of work focused on the design of relay systems using SC and finite-alphabet signals. Optimal precoders for an AF half-duplex relay channel were designed to obtain the maximum coding gain in \cite{b6}. In this work, the source had two signal vectors to transmit in each transmission block while the relay had no information of its own to transmit. During the first time slot, the source transmitted its first signal vector to both the destination and relay. In the second time slot, the source transmitted its second signal vector to the destination while the relay simply amplified and forwarded what it has received from the first time slot to the destination. The two signal vectors transmitted in the second time slot were considered to be received in a superposition manner at the destination. In \cite{b7}, an optimal superposition-coded relay scheme and a sub-optimal switched-power SC scheme were proposed by maximizing the equivalent squared minimum distance (ESMD) that determined the error performance. Moreover, the error performance of a high-rate bit-interleaved coded modulation with iterative decoding (BICM-ID) system using superposition modulation over a non-orthogonal amplify-and-forward (NAF) half-duplex single relay channel was studied in \cite{b8}. Although the above studies have made great efforts on the SC-based relay systems, only the source was considered to have information to transmit while the relay was only responsible for forwarding the source information. However, in many practical systems (e.g., wireless sensor networks), the relay can be a normal network node that has its own information to transmit while helping forward the source's signal.  As far as we know, there is no existing work that designs SC-based relay systems with both the source and relay having their own information to transmit using finite-alphabet constellations.

Motivated by this gap, in this paper we concentrate on the signal design in a Gaussian AF half-duplex relay channel with SC being applied at the relay, which allows the source and relay to transmit their own information within two time slots. In our design, the relay superimposes its own information on the amplified version of the received noisy signal from the source before forwarding the superposed signal to the destination. For practical consideration, QAM constellations of not necessarily the same order are assumed to be adopted at the source and relay. We optimize the \textit{weight coefficients} of both the source and relay to maximize the minimum Euclidean distance of the composite constellation of the received signal at the destination, which dominates the error performance of the considered system in high signal-to-noise ratio (SNR) regimes. The formulated problem is a \textit{mixed continuous-discrete} problem that is in general non-trivial to resolve. To tackle it, we resort to the \textit{punched Farey sequence} framework developed in \cite{b9} to strategically segment the entire feasible region of the formulated optimization problem into a finite number of sub-intervals and obtain the closed-form optimal solution of the formulated problem in each sub-interval. We then acquire the optimal solution to the formulated problem by comparing the solutions on each sub-interval. Simulation results demonstrate that our proposed SC-based scheme can achieve better average bit error rate (BER) performance than that using conventional time division multiple access (TDMA). Furthermore, the performance gain increases as the channel gain of the relay-destination link increases. 

\section{System Model}
We consider a Gaussian AF relay channel with SC being applied at the relay as depicted in Fig.\ 1, where both the source and relay have their own information to transmit to the destination. In our considered system, we assume no direct link between the source and destination. During each transmission block, it is assumed that two time slots with equal length are used. In the first time slot, the source transmits its information to the relay. The received signal at the relay can be given by 
\begin{equation} 
z_{1} = h_{1}(v_{1}x_{1})+\xi_{1},
\end{equation}
\noindent where $h_1$ denotes the complex channel coefficient of the source-relay link, $v_1$ is the weight coefficient adjusting the transmitted signal at the source, and $\xi_{1}$ is the additive zero-mean, circularly symmetric complex Gaussian (CSCG) noise with variance $2\sigma_{1}^{2}$, i.e., $\xi_{1} \sim \mathcal{CN}\left(0,2\sigma_{1}^{2}\right)$.

In the second time slot, the relay amplifies the signal received from the source, then superimposes its own signal and forwards the superposed signal to the destination. The received signal at the destination can thus be expressed as 
\begin{equation} 
z_{2} \!=\! h_{2}(v_{2}x_{2}+v_{3}z_{1})+\xi_{2} \!=\! h_{2}(v_{2}x_{2}+v_{3}h_{1}v_{1}x_{1}+v_{3}\xi_{1})+\xi_{2},	
\end{equation}
\noindent where $h_{2}$ is the complex channel coefficient of the relay-destination link, $v_2$ and $v_3$ are the weight coefficients adjusting the superposed signal at the relay, and $\xi_{2} \sim \mathcal{CN}\left(0,2\sigma_{2}^{2}\right)$ is the additive zero-mean CSCG noise on this link. It is assumed that the source knows perfect channel state information (CSI) of the source-relay link and the relay knows perfect CSI of the relay-destination link. The transmitted symbols $x_{k}$ are chosen randomly, independently and equally likely from the $M_{k}^{2}$-ary square QAM constellation $\mathcal{Q}_{k}$. Based on the knowledge of perfect CSI, we assume that $v_1$ completely offsets the phase rotation brought by $h_1$ to make sure there is no phase difference between the transmitted signals in $z_1$ and $x_2$, and thus they can be superimposed at the relay by means of SC. Similarly, $v_2$ and $v_3$ are also assumed to completely offset the phase rotation brought by $h_2$.

Although (1), (2) are in the form of complex baseband, the transmitting oscillators at the source and relay can only generate real sinusoids rather than complex signals. 
\begin{figure}[tbp]
	\centerline{\includegraphics[height=2.0cm]{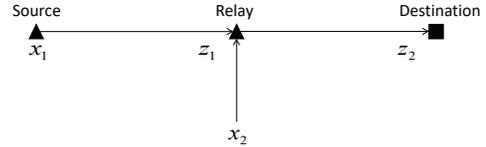}}
	\vspace{-2mm}
	\caption{Gaussian AF Relay Channel}
	\vspace{-6mm}
	\label{fig:two-hoop-relay-model-1}
\end{figure}
This means that the considered two-dimensional QAM constellation actually could be split into two one-dimensional pulse amplitude modulation (PAM) constellations, which are called in-phase and quadrature branches \cite{b10}, to be transmitted. Consequently, the complex Gaussian AF relay channel of the relay-destination link given in (2) can be decomposed into two real-scalar Gaussian AF relay channels. In this paper, we will consider the symbols whose in-phase and quadrature branches are symmetric, i.e., equal power allocation between the two branches. However, it is worth mentioning that our work can be readily extended to the unsymmetric case \cite{b11}. Because of the symmetry of the two sub-channels, we simply need to design one of the two real-scalar Gaussian AF relay channels with PAM constellation sets. Designing the constellation of the in-phase real-scalar Gaussian AF relay channel will be considered hereafter.  The design of the quadrature branch is similar and thus is omitted for brevity.

Mathematically, we can rewrite (2) as \vspace{-1mm}
\begin{equation}
\begin{aligned}
z_{2}=&\ |{h_2}||{h_1}|\exp (j\arg ({h_2})){v_3}\exp (j\arg ({h_1})){v_1}{x_1}\\ 
&\,+|{h_2}|\exp (j\arg ({h_2})){v_2}{x_2}\\
&\,+|{h_2}|\exp (j\arg ({h_2})){v_3}{\xi _1}+{\xi _2}.
\end{aligned} \vspace{-1.5mm}
\end{equation}
	
\noindent For the sake of notations, we assume that ${y_2} = {\mathop{\rm Re}\nolimits} ({z_2})$, ${w_1}{s_1} = {\mathop{\rm Re}\nolimits} ({v_1}{x_1}\exp (j\arg ({h_1})))$, ${w_2}{s_2} = {\mathop{\rm Re}\nolimits} ({v_2}{x_2}\exp (j\arg ({h_2})))$, ${w_3} = {\mathop{\rm Re}\nolimits} ({v_3}\exp (j\arg ({h_2})))$, ${n_1} = {\mathop{\rm Re}\nolimits} ({\xi _1})$ and ${n_2} = {\mathop{\rm Re}\nolimits} ({\xi _2})$, where ${\mathop{\rm Re}\nolimits} ( \cdot )$ is the real part of a complex number. Besides,  $w_1$, $w_2$ and $w_3$
are real non-negative weight coefficients controlling the minimum Euclidean distance of the PAM constellation sets during the actual transmission. Note that we do not need information-bearing symbol $s_3$ to replace ${v_3}\exp (j\arg ({h_2}))$ because it does not contain any transmitted symbols $x_k$. Now, the in-phase branch of (3) can be expressed as \vspace{-1mm}
\begin{equation}
\begin{aligned}
y_2=&\ |h_2||h_1|\mathrm{Re}(v_3\exp(jarg(h_2))v_1x_1\exp(jarg(h_1)))\\
&\,+|h_2|\mathrm{Re}(v_2x_2\exp(jarg(h_2))\\	&\,+|h_2|\mathrm{Re}(v_3\exp(jarg(h_2))\xi_1)+\mathrm{Re}(\xi_{2}).
\end{aligned} \vspace{-1.5mm}
\end{equation}

From the given assumption that $v_3$ completely offsets the phase rotation brought by $h_2$, we can write that ${v_3} = |{v_3}|\exp (j\arg ( - {h_2}))$, and ${v_3}\exp (j\arg ({h_2}))$ is thus a real non-negative scalar itself, i.e., ${w_3} = {\mathop{\rm Re}\nolimits} ({v_3}\exp (j\arg ({h_2}))) = {v_3}\exp (j\arg ({h_2}))$. As a result, (4) can be rewritten as 
\begin{equation} 
\begin{aligned} 
y_2 &= |h_2||h_1|w_3\mathrm{Re}(v_1x_1\exp(jarg(h_1)))+|h_2|w_2s_2\\
&\,\quad+|h_2|w_3\mathrm{Re}(\xi_1)+n_2\\
&= |h_2||h_1|w_3w_1s_1+|h_2|w_2s_2+|h_2|w_3n_1+n_2,\\ 
\end{aligned}
\end{equation}	

\noindent where $ n_1 \sim \mathcal{N}\left(0,\sigma_{1}^{2}\right)$ and $ n_2 \sim \mathcal{N}\left(0,\sigma_{2}^{2}\right)$ are independent and identically distributed (i.i.d.) real additive white Gaussian components because of the CSCG noise terms $\xi_{1}$ and $\xi_{2}$. As a guarantee of generality, we assume that ${v_1}{x_1}\exp (j\arg ({h_1})) \in \mathcal{Q}_{1}$ and ${v_2}{x_2}\exp (j\arg ({h_2})) \in \mathcal{Q}_{2}$, where $\mathcal{Q}_{1} $ and $\mathcal{Q}_{2} $ are $M_{1}^{2}$ and $M_{2}^{2}$-ary square QAM constellations given by $\mathcal{Q}_{1} \triangleq \left\{ \pm w_{1}(2k-1) \pm w_{1}(2l-1)j: k, l = 1,...,M_{1}/2 \right\}$ and $\mathcal{Q}_{2} \triangleq \left\{ \pm w_{2}(2k-1) \pm w_{2}(2l-1)j: k, l = 1,...,M_{2}/2 \right\}$ respectively. Therefore, the information-bearing symbols $s_{1} \in  \left\{ \pm(2k-1) \right\}_{k=1}^{M_{1}/2}$ and $s_{2} \in  \left\{ \pm(2k-1) \right\}_{k=1}^{M_{2}/2}$ are extracted from the standard PAM constellations with equal probability.

Consider that the individual average power constraints of the source and relay are $P_1$ and $P_2$ respectively. We then have the following constraints, 
\begin{subequations}
	\begin{equation}
	\mathbb{E}[|v_{1}x_{1}|^{2}] \leq P_{1},
	\end{equation} 
	\begin{equation} \mathbb{E}[|v_{3}h_{1}v_{1}x_{1}|^{2}]+\mathbb{E}[|v_{2}x_{2}|^{2}]+2\sigma_{1}^{2}|v_{3}|^{2} \leq P_{2}.  
	\end{equation} 
\end{subequations}

Recall that the symbols considered in this paper have equal power between their in-phase and quadrature branches. We thus have $\mathbb{E}[|{v_1}{x_1}{|^2}] = 2\mathbb{E}[|{w_1}{s_1}{|^2}] \le {P_1}$, i.e., the power constraint of the in-phase signal transmitted by the source should be 
\begin{equation}
\mathbb{E}[|w_{1}s_{1}|^{2}] \leq \frac{1}{2}P_1.
\end{equation}

For the terms on the left-hand side of (6b), we first have $\mathbb{E}[|v_3h_1v_1x_1|^2] = |{h_1}{|^2}\mathbb{E}[||{v_3}|\exp (j\arg ( - {h_2})){v_1}{x_1}\exp (j\arg ({h_1}))|^2]
= |h_1|^2|v_3|^2\mathbb{E}[|v_1x_1\exp(j\arg(h_1))|^2] = 2|{h_1}{|^2}w_3^2\mathbb{E}[|{w_1}{s_1}{|^2}]$ by noting $v_3=|{v_3}|\exp (j\arg ( - {h_2}))$. Considering that $w_1$ and $w_3$ are independent, we further have $\mathbb{E}[|v_3h_1v_1x_1|^2] = 2\mathbb{E}[||h_1|w_3w_1s_1|^2]$. Similarly, the second and third items of (6b) can be derived as $\mathbb{E}[|v_2x_2|^2] = 2\mathbb{E}[|w_2s_2|^2]$ and $|v_3|^2\sigma_1^2 = w_3^2\sigma_1^2 = \mathbb{E}[|w_3n_1|^2]$, respectively. Based on these derivations, we can rewrite (6b) as $\mathbb{E}[|v_{3}h_{1}v_{1}x_{1}|^{2}]+\mathbb{E}[|v_{2}x_{2}|^{2}]+2\sigma_{1}^{2}|v_{3}|^{2} = 2\mathbb{E}[||{h_1}|{w_3}{w_1}{s_1}{|^2}] + 2\mathbb{E}[|{w_2}{s_2}{|^2}] + 2\mathbb{E}[|{w_3}{n_1}{|^2}] \leq P_{2}.$ Therefore, the power constraint of the in-phase signal transmitted by the relay can be expressed as 
\begin{equation}
\mathbb{E}[||h_1|w_3w_1s_1|^2]+\mathbb{E}[|w_2s_2|^2]+\mathbb{E}[|w_3n_1|^2] \leq \frac{1}{2}P_2. 
\end{equation}

Jointly optimizing the weight coefficients $w_1$, $w_2$ and $w_3$ to minimize the average error probability of the received signals is the focus of this paper. We will formally formulate and resolve this optimization problem in the following section for any given QAM constellation sizes of both transmitted signals, subject to the individual average power constraints of both the source and relay.

\section{The Weight Coefficients Design for SC in Real-Scalar Gaussian AF Relay Channel}

In this section, we will optimize the weight coefficients $w_1$, $w_2$ and $w_3$ to minimize the error performance of the system. As the in-phase and quadrature branches are symmetric, the corresponding optimization problem of the quadrature component can be solved using exactly the same method and hence is omitted for brevity.

\subsection{Problem Formulation}

In our scheme, the transmitted signals from the source and relay are superimposed together using SC at the relay before they are transmitted to the destination. Furthermore, we adopt joint decoding by applying a coherent joint maximum likelihood (ML) detector at the destination to estimate the received signals in a symbol-by-symbol fashion. At the destination, the estimated signals can be expressed as 
\begin{equation}
(\hat{s_1}, \hat{s_2}) = \arg \mathop {\min }\limits_{({s_1},{s_2})} |{y_2} - (|{h_2}||{h_1}|{w_3}{w_1}{s_1} + |{h_2}|{w_2}{s_2})|. 
\end{equation}

According to the nearest neighbor approximation method for the ML receiver \cite[Ch. 6.1.4]{b10}, we know that the average error rate is dominated by the minimum Euclidean distance of the received constellation points at high SNRs. As a result, in this paper, we intend to work out the optimal value of $\left(w_{1},w_{2},w_{3}\right)$ to maximize the minimum Euclidean distance of the received constellation points so that the error rate is reduced in essence. For given pairs of in-phase signals $\left(s_{1},s_{2}\right)$ and $\left({\tilde {s}}_{1},{\tilde {s}}_{2}\right)$, the normalized Euclidean distance between the two received signals $y_2\left(s_{1},s_{2}\right)$ and $y_2\left({\tilde {s}}_{1},{\tilde {s}}_{2}\right)$ at the destination in the noise free case is given by 
\begin{equation}
\begin{aligned}
&\qquad\qquad\ \  \frac{{|y_{2}(s_{1},s_{2})-y_{2}({\tilde {s}}_{1},{\tilde {s}}_{2})|}}{{\sqrt {|{h_2}{|^2}w_3^2\sigma _1^2 + \sigma _2^2} }} \\
&=\frac{{||{h_2}| \cdot |{h_1}|{w_3}{w_1}({s_1} - \tilde{s}_1) - |{h_2}|{w_2}(\tilde{s}_2 - {s_2})|}}{{\sqrt {|{h_2}{|^2}w_3^2\sigma _1^2 + \sigma _2^2} }}. 
\end{aligned} 
\end{equation}

As the information-bearing symbols $s_{1}, {{\tilde{s}}_{1}}, s_{2},$ and ${{\tilde{s}}_{2}}$ are drawn from the standard PAM constellations as mentioned above, all of them are odd numbers. Hence, we can write that $s_{1}-{\tilde{s}}_{1} = 2n$  and ${\tilde{s}}_{2}-s_{2} = 2m$, where $n \in \mathbb{Z}_{M_{1}-1} $ and $m \in \mathbb{Z}_{M_{2}-1}$ with $\mathbb{Z}_{N} \triangleq \left\{0,\pm 1, ..., \pm N \right\}$. In addition, $\mathbb{Z}^{2}_{(M_{1}-1,M_{2}-1)} \triangleq \left\{(a,b) : a \in \mathbb{Z}_{M_{1}-1}, b \in \mathbb{Z}_{M_{2}-1} \right\}$ is also defined here. According to the above definitions, we can deduce that $(m,n) \ne (0,0)$ $\left(\mathrm{i.e.}, m \neq 0\ \mathrm{or}\ n \neq 0\right)$ from $\left(s_{1}, s_{2}\right) \neq \left({\tilde{s}}_{1},{\tilde{s}}_{2}\right)$. To proceed, we define 
\begin{equation}
\begin{aligned}
d(m,n) &=\frac{1}{2}\frac{|y_{2}(s_{1},s_{2})-y_{2}({\tilde {s}}_{1},{\tilde {s}}_{2})|}{\sqrt{|h_2|^2w_3^2\sigma_{1}^2+\sigma_{2}^2}}\\
&= \frac{{|{h_2}| \cdot ||{h_1}|{w_3}{w_1}n - {w_2}m|}}{{\sqrt {|{h_2}{|^2}w_3^2\sigma _1^2 + \sigma _2^2} }},
\end{aligned} 
\end{equation}

\noindent where $(m,n) \in \mathbb{Z}^{2}_{(M_{1}-1,M_{2}-1) \backslash {(0,0)}}$ and $\mathcal{A} \backslash \mathcal{B} \buildrel \Delta \over =  \{ x \in \mathcal{A}\ \mathrm{and} \  x \notin \mathcal{B}\} $. In the following, the formal formulation of the max-min optimization problem is described.

\textit{\underline{Problem} 1: Optimal design of weight coefficients for SC in real-scalar AF relay channel with PAM constellation:} Devise the optimal value of $\left({w}_{1}^{*}, {w}_{2}^{*}, {w}_{3}^{*}\right)$ which is subject to the power constraints in order to maximize the minimum Euclidean distance $d(m,n)$ of the received signal constellation points, i.e., 
\begin{small}
\begin{subequations}
	\begin{equation}
	(w_1^*, w_2^*, w_3^*) \!=\! \arg \mathop {\max }\limits_{({w_1},{w_2},{w_3})} \mathop {\min }\limits_{(m,n) \in \mathbb{Z}_{({M_1} \!-\! 1,{M_2} \!-\! 1)}^2\backslash \ (0,0) } d(m,n) 
	\end{equation}	
	\begin{equation} 
	\begin{aligned} 
	\mathbf{s.t.} \quad   &0 < w_1^2 \le \frac{{3{P_1}}}{{2(M_1^2 - 1)}}\quad \\
	&\mathrm{and} \quad  0 < \frac{|h_1|^2(M_1^2-1)}{3}w_1^2w_3^2 \\
	&\qquad\qquad\qquad+ \frac{(M_2^2-1)}{3}w_2^2 + w_3^2\sigma_{1}^2 \leq \frac{P_{2}}{2},
	\end{aligned} 
	\end{equation} 
\end{subequations}
\end{small}\noindent where (12b) is the power constraint derived from (7) and (8).

We can see that the inner optimization problem is on the discrete variables $m$ and $n$, while the outer one is determined by the continuous ones $\left(w_1, w_2, w_3\right)$. This means that Problem 1 is a mixed continuous-discrete optimization problem which is generally difficult to solve. As far as we know, only a handful of open literature, such as \cite{b12},  \cite{b13}, work on solving this kind of optimization problem while all of them only provide numerical solutions. In this paper, we apply the punched Farey sequence framework, which is developed from the original \textit{Farey sequence} applied in \cite{b14}, to methodically and optimally solve the formulated problem. By taking advantage of the punched Farey sequence, the entire feasible region of $\left(w_1, w_2, w_3\right)$ can be divided into a finite number of mutually exclusive sub-intervals. Within each of these sub-intervals, we achieve the closed-form optimal solution to the formulated problem. By doing this, we can acquire the overall maximum value of Problem 1 by taking the maximum value of the inner function among all the sub-intervals. We now first deal with the inner optimization problem in (12) given by:

\textit{\underline{Problem} 2: Finding differential pairs with the minimum Euclidean distance:} 
\begin{equation}
\begin{aligned}
&\mathop {\min }\limits_{(m,n) \in \mathbb{Z}_{({M_1} - 1,{M_2} - 1)}^2\backslash \ (0,0) } d(m,n)\\
= &\mathop {\min }\limits_{(m,n) \in \mathbb{Z}_{({M_1} - 1,{M_2} - 1)}^2\backslash \ (0,0) } \frac{{|{h_2}| \cdot ||{h_1}|{w_3}{w_1}n - {w_2}m|}}{{\sqrt {|{h_2}{|^2}w_3^2\sigma _1^2 + \sigma _2^2} }}. 
\end{aligned} 
\end{equation} 

Actually, we can observe that the essential problem to be solved in Problem 2 is to find rational approximations $\left(m,n\right)$ of the irrational numbers determining the minimum Euclidean distance, where the punched Farey sequence can be of great help. Therefore, in the following section, we will give the solution to Problem 2 by means of the punched Farey sequence. It is also worth pointing out that finding the closed-form solution to the optimal differential pairs $\left(m,n\right)$ is tricky because the channel coefficient $|h_1|$ which is crucial to the optimal solution, can vary across the whole positive real axis. Furthermore, the optimal weight coefficients $\left(w_1, w_2, w_3\right)$ cannot be determined beforehand for the inner optimization problem.\\

\noindent \textit{B. The Minimum Euclidean Distance of the Constellation Points of the Received Signal}

In this section, we aim to solve Problem 2 to find the differential pairs $\left(m,n\right)$ having the minimum Euclidean distance. For this purpose, we first introduce the definition of the punched Farey sequence developed in \cite{b9}.

\textit{\underline{Definition} 1: Punched Farey sequence:} The punched Farey sequence $\mathfrak{P}_K^L$ is the ascending sequence of irreducible fractions whose denominators are no greater than $K$ and numerators are no greater than $L$.

According to this definition, the punched Farey sequence $\mathfrak{P}_K^L = \left(\frac{{{b_k}}}{{{a_k}}}\right)_{k = 1}^{\left| \mathfrak{P}_{K}^{L}\right|}$ is a sequence of irreducible fractions with $0  \le {a_k} \le K$, $0  \le {b_k} \le L$ and $\langle {a_k},{b_k}\rangle  = 1$ arranged in an increasing order, where $\langle {a_k},{b_k}\rangle$ denotes the greatest common divisor of non-negative integers $a_k, b_k$. In addition, $\left| \mathfrak{P}_{K}^{L} \right|$ denotes the cardinality of $\mathfrak{P}_K^L$. An example of the punched Farey sequence is given as follows:

\textit{\underline{Example} 1:} $\mathfrak{P}_4^3$ is the ordered sequence $\left(\frac{0}{1},\frac{1}{4},\frac{1}{3},\frac{1}{2},\frac{2}{3},\frac{3}{4},\frac{1}{1},\frac{3}{2},\frac{2}{1},\frac{3}{1}, \frac{1}{0}\right)$.

Some elementary properties, which are helpful to find the differential pairs $(m,n)$ having the minimum Euclidean distance, are presented with proof in \cite{b9}. Using these properties, the solution to Problem 2 can be given in the following proposition.

\textit{\underline{Proposition} 1:} For any $\frac{n_1}{m_1}$, $\frac{n_2}{m_2}$, $\frac{n_3}{m_3}$, $\frac{n_4}{m_4} \in \mathfrak{P}_{{M_2} - 1}^{{M_1}-1}$ with 	$\left| \mathfrak{P}_{{M_2} - 1}^{{M_1} - 1} \right| \ge 4$, such that $\frac{{{n_1}}}{{{m_1}}} < \frac{{{n_2}}}{{{m_2}}} < \frac{{{n_3}}}{{{m_3}}} < \frac{{{n_4}}}{{{m_4}}}$ and $\frac{n_2}{m_2}$, $\frac{n_3}{m_3}$ are successive in $\mathfrak{P}_{{M_2} - 1}^{{M_1}-1}$, we have 1) If $\frac{w_2}{|h_1|w_3w_1} \in \left(\frac{{{n_2}}}{{{m_2}}},\frac{{{n_2} + {n_3}}}{{{m_2} + {m_3}}}\right)$, then ${\min _{(m,n) \in \mathbb{Z}_{({M_1} - 1,{M_2} - 1)}^2\backslash \ (0,0)}}d(m,n) = d({m_2},{n_2}) = \frac{{|{h_2}|}}{{\sqrt {|{h_2}{|^2}w_3^2\sigma _1^2 + \sigma _2^2} }}({w_2}{m_2} - |{h_1}|{w_3}{w_1}{n_2})$; 2) If $\frac{w_2}{|h_1|w_3w_1} \in \left(\frac{n_2+n_3}{m_2+m_3},\frac{n_3}{m_3}\right)$, then ${\min _{(m,n) \in \mathbb{Z}_{({M_1} - 1,{M_2} - 1)}^2\backslash \ (0,0)}}d(m, n) = d({m_3},{n_3}) = \frac{{|{h_2}|}}{{\sqrt {|{h_2}{|^2}w_3^2\sigma _1^2 + \sigma _2^2} }}(|{h_1}|{w_3}{w_1}{n_3}-{w_2}{m_3})$.

The proof is omitted due to space limitation.\\

\noindent \textit{C. Optimal Solution to Problem 1}

With the optimal solution to the inner problem given in Proposition 1, we are now ready to find the optimal solution to the outer optimization on $\left(w_1,w_2,w_3\right)$ of Problem 1. We observe that $w_1$ has a separate power constraint, while $w_2$ and $w_3$ have a joint power constraint. We thus first find the optimal solution of $w_1$, which is given in the following lemma:

\textit{\underline{Lemma} 1:} The optimal value of $w_1$ within its average power constraint to maximize the minimum Euclidean distance of $d(m,n)$ given in (13) should be chosen as:
\begin{equation}
w_1 = \sqrt {\frac{{3{P_1}}}{{2(M_1^2 - 1)}}}.
\end{equation}

The proof is omitted due to space limitation.

To proceed, we first write the entire feasible region of Problem 1 in terms of $\left(w_2, w_3\right)$ as  $\mathcal {U}=\Big\{(w_2,w_3): 0 < |{h_1}{|^2}{P_1}w_3^2 + \frac{{M_2^2 - 1}}{3}w_2^2 + w_3^2\sigma _1^2 \le \frac{{{P_2}}}{2}\Big\}$. According to the definition of the punched Farey sequence, we know that the punched Farey sequence $\mathfrak{P}_{{M_2} - 1}^{M_1-1}=\left(\frac{{{b_1}}}{{{a_1}}},\frac{{{b_2}}}{{{a_2}}},...,\frac{{{b_C}}}{{{a_C}}}\right)$, where $C=\left| \mathfrak{P}_{{M_2} - 1}^{{M_1} - 1} \right|$, can divide the positive real axis into $C-1$ mutually exclusive sub-intervals. By denoting $\mathcal{A}_k=\Big\{ (w_2,w_3): 0 < |{h_1}{|^2}{P_1}w_3^2 + \frac{{M_2^2 - 1}}{3}w_2^2 + w_3^2\sigma _1^2 \le \frac{{{P_2}}}{2}, \frac{{{b_k}}}{{{a_k}}} < \frac{{{w_2}}}{{{w_3}}}\sqrt {\frac{{2(M_1^2 - 1)}}{{3|{h_1}{|^2}{P_1}}}}  \le \frac{{{b_{k + 1}}}}{{{a_{k + 1}}}}\Big\}$, $k=1, 2, ..., C-1$ where $\mathcal{U}= \bigcup\nolimits_{k = 1}^{C - 1} {{\mathcal{A}_k}}$, we can solve Problem 1 by restricting $(w_2, w_3) \in {\mathcal{A}_k}$. Concretely speaking, we intend to find the optimal $\left(w_2^*(k), w_3^*(k)\right)$ such that
\begin{subequations}
	\begin{equation}
	g\left(\frac{{{b_k}}}{{{a_k}}},\frac{{{b_{k + 1}}}}{{{a_{k + 1}}}}\right) =\! \mathop {\max }\limits_{({w_2},{w_3})} \mathop {\min }\limits_{(m,n) \in \mathbb{Z}_{({M_1} - 1,{M_2} - 1)}^2\backslash \ (0,0) } d(m,n)
	\end{equation}
	\begin{equation}
	\begin{aligned}
	\mathbf{s.t.}  \quad  &\frac{{{b_k}}}{{{a_k}}} \!<\! \frac{{{w_2}}}{{{w_3}}}\sqrt {\frac{{2(M_1^2 - 1)}}{{3|{h_1}{|^2}{P_1}}}}  \!\le\! \frac{{{b_{k + 1}}}}{{{a_{k + 1}}}} \quad \\
	&\mathrm {and} \quad 0 \!<\! \left(|{h_1}{|^2}{P_1} \!+\! \sigma _1^2\right)w_3^2 \!+\! \frac{{M_2^2 - 1}}{3}w_2^2 \!\le\! \frac{{{P_2}}}{2}.
	\end{aligned}
	\end{equation}
\end{subequations}
	
By applying the results obtained in Proposition 1, we can acquire the optimal solution to problem (15) given by the following theorem.

\textit{\underline{Theorem} 1:} The optimal solution to (15) is given as follows:
\begin{equation}
\begin{aligned}
&g(\frac{{{b_k}}}{{{a_k}}},\frac{{{b_{k \!+\! 1}}}}{{{a_{k \!+\! 1}}}}) \!=\! \sqrt {\frac{{3|{h_1}{|^2}|{h_2}{|^2}{P_1}{P_2}}}{{X\sigma _2^2{{({a_k} \!+\! {a_{k \!+\! 1}})}^2} \!+\! Y\sigma _2^2{{({b_k} \!+\! {b_{k \!+\! 1}})}^2} \!+\! Z\sigma _1^2}}},\\
&\mathrm {with}\ (w_2^*(k),w_3^*(k)) \!=\! \!\Bigg(\!\sqrt{\frac{3|h_1|^2P_1P_2}{X(\frac{a_k\!+\!a_{k\!+\!1}}{b_k\!+\!b_{k\!+\!1}})^2\!+\!Y}},\\
&\,\,\quad\,\qquad\qquad\qquad\qquad\ \;\sqrt {\frac{{2(M_1^2 \!-\! 1)}}{{3|{h_1}{|^2}{P_1}}}} \left(\frac{{{a_k} \!+\! {a_{k \!+\! 1}}}}{{{b_k} \!+\! {b_{k \!+\! 1}}}}\right)w_2^*(k)\!\Bigg)\!,
\end{aligned}
\end{equation}

\noindent where $X = 4(|h_1|^2P_1+\sigma_1^2)(M_1^2-1)$, $Y = 2|h_1|^2P_1(M_2^2-1)$ and $Z = 2|h_2|^2P_2(M_1^2-1)$. 

The proof is omitted due to space limitation.

\textit{\underline{Summary}:} Based on Theorem 1, we can find $g\left(\frac{{{b_k}}}{{{a_k}}},\frac{{{b_{k + 1}}}}{{{a_{k + 1}}}}\right)$ for $k=1, 2, ..., C-1$, which means the maximum value of the minimum Euclidean distance on each sub-interval determined by the punched Farey sequence can be obtained by simple algebraic operations. By comparing all these $C-1$ values, we can readily acquire the sub-interval where the overall maximum value of Problem 1 appears, and we finally find the optimal value of $(w_2, w_3)$.

Note that the computational complexity of our proposed method is just $\mathcal{O}(C)$.
In the absence of our method, the exhaustive search approach is commonly used to solve Problem 1 with computational complexity as high as $\mathcal{O}(D_1D_2M_1M_2)$, where $D_1$ and $D_2$ are the discretization levels for the feasible ranges of $w_2$ and $w_3$ respectively. Obviously, our proposed method greatly reduces the computational complexity.

\section{Simulation Results and Discussions}
In this section, computer simulations are carried out to demonstrate the BER performance of our proposed SC-based design compared with that using TDMA for the transmission of the source and relay information. Without loss of generality, we assume that all channels are subject to Rayleigh fading such that $h_k \sim \mathcal{CN}\left(0,2\delta _k^2\right)$, $k=1,2$. During each transmission block, the source and relay each use one time slot to transmit information in our proposed SC-based scheme. In contrast, in TDMA-based scheme, the relay uses two time slots to separately forward the source's information and transmit its own information, and the source still uses one. In terms of the transmission power, we assume that the source transmits at a power value of $\frac{2}{3}$ and 1 in SC-based scheme and TDMA-based scheme respectively $\left(\mathrm{i.e.}, {P_{\mathrm{source}\_\mathrm{SC}}} = \frac{2}{3},  {P_{\mathrm{source}\_\mathrm{TDMA}}}=1\right)$ while the instantaneous transmission power of the relay is $\frac{4}{3}$ and 1 in SC-based scheme and TDMA-based scheme respectively $\left(\mathrm{i.e.}, {P_{\mathrm{relay}\_\mathrm{SC}}} = \frac{4}{3},  {P_{\mathrm{relay}\_\mathrm{TDMA}}}=1\right)$. These power constraint numbers are justified as follows: If the instantaneous transmission power of the relay in each time slot is assumed to be 1 in TDMA-based scheme, where a total of three time slots are used per transmission block, its average transmission power is equal to $\frac{2}{3}$. Therefore, in SC-based scheme, the relay should transmit with the power of $\frac{4}{3}$, which results in an average power of $\frac{2}{3}$, to ensure the fairness of transmission power. The instantaneous transmission power of the source can be determined in the same way. 

For the sake of comparison, we assume that, the number of transmission blocks of the TDMA-based scheme is two thirds of that transmitted in SC-based scheme over the same time period. Therefore, the constellation sizes of the TDMA-based scheme should be increased to $M_{1}^{\frac{3}{2}}$- and $M_{2}^{\frac{3}{2}}$-ary PAM constellations (i.e., $M_{1}^{3}$- and $M_{2}^{3}$-ary QAM constellations) to maintain the same data rate. Note that we should select values for $M_1$ and $M_2$ carefully because the constellation sizes in TDMA-based scheme are special.

We first study the average BER of both schemes against the SNR. As depicted in Fig.\ 2, we have the comparison in two cases, where the variances of the channel coefficients are $\left(\delta _1^2,\delta _2^2\right) = \left(1,2\right)$ and $\left(\delta _1^2,\delta _2^2\right) = \left(1,3\right)$ respectively. In both cases, we assume that the source adopts 256-QAM and the relay adopts 16-QAM in our proposed SC-based design. Correspondingly, 4096-QAM and 64-QAM should be adopted by the source and relay, respectively, in TDMA-based scheme. In both simulated cases, our proposed SC-based design outperforms the TDMA-based scheme in moderate and high SNR regimes. Specifically, the SNR gain of our proposed scheme over the TDMA-based scheme is approximately 2dB at the BER of $10^{-3}$ when $\left(\delta _1^2,\delta _2^2\right) = \left(1,2\right)$, while the performance gain is about 4dB when $\left(\delta _1^2,\delta _2^2\right) = \left(1,3\right)$. We note that the performance gain increases as the channel gain of the relay-destination link increases. 

As can be seen from Fig.\ 2, the BER performance gain of our proposed SC-based scheme is related to the relative channel gains of both links. Inspired by this observation, we now study the BER against the relative strength of the channels by fixing the variance of the source-relay link as $\delta_1^2=1$ and changing the variance of the relay-destination link, i.e., $\delta_2^2$, which is represented in dB. As depicted in Fig.\ 3, in this simulation, we also consider two cases with the SNRs being 40dB and 50dB respectively. In this first case of SNR = 40dB, it is apparent that when $\delta_2^2$ is larger than 0dB (i.e., the variance of the relay-destination link is larger than 1), the BER performance of the SC-based scheme becomes better than that of the TDMA-based scheme, but the BER gain gradually reaches saturation. This is mainly because the error performance of both schemes is dominated by the source-relay link when the 
\begin{figure}[tbp] \vspace{-2mm}
	\centerline{
		\includegraphics[width=0.9\linewidth]{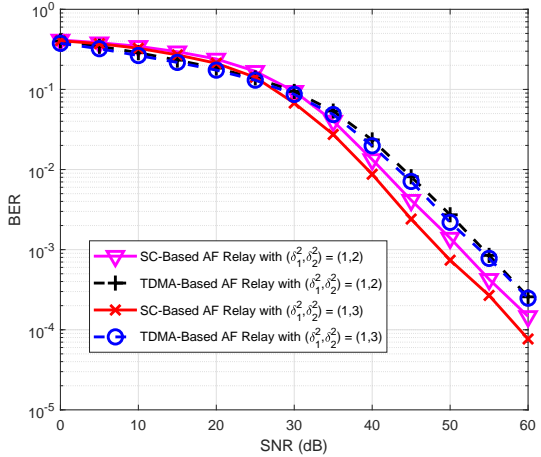}
	} \vspace{-4mm}
	\caption{Comparison between the proposed SC-based AF relay and TDMA-based AF relay in different channel conditions, where 256-QAM and 16-QAM are used in proposed SC-based AF relay while 4096-QAM and 64-QAM are used in TDMA-based relay.}
	\label{fig_2} 
\end{figure} 
\begin{figure}[tbp] \vspace{-5mm}
	\centerline{
		\includegraphics[width=0.9\linewidth]{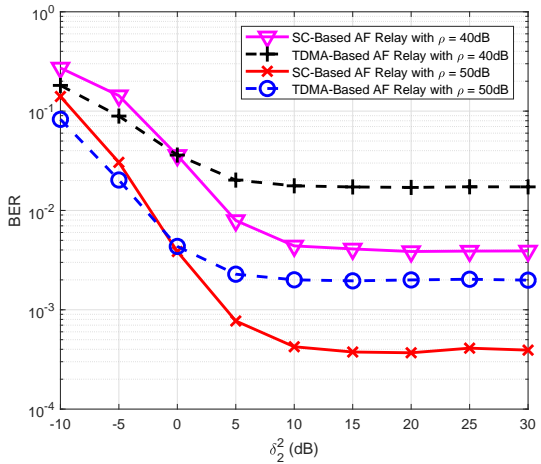}
	} \vspace{-2.5mm}
	\caption{Comparison between the proposed SC-based AF relay and TDMA-based AF relay under different SNRs, where 256-QAM and 16-QAM are used in proposed SC-based AF relay while 4096-QAM and 64-QAM are used in TDMA-based relay.}
	\label{fig_3} \vspace{-4.7mm}
\end{figure}relay-destination link becomes rather good. Since the source in our proposed SC-based scheme uses an instantaneous power value of $\frac{2}{3}$ to transmit with 256-QAM while that in TDMA-based scheme transmits with 4096-QAM using an instantaneous power value of 1, the minimum Euclidean distance between the constellation points in our scheme is much larger than that in TDMA-based scheme, which makes the proposed scheme essentially outperform the TDMA-based scheme in BER performance. In the second case of SNR = 50dB, the BER gain of our proposed SC-based scheme becomes even more significant, which is consistent with what we observed in Fig.\ 2.

\section{Conclusions}

In this paper, we developed a practical design framework for the Gaussian amplify-and-forward (AF) relay channel with superposition coding (SC) being applied at the relay, which allowed the source and relay to transmit their own information using practical quadrature amplitude modulation (QAM)  constellations within two time slots. Specifically, to improve the error performance, we formulated a \textit{mixed continuous-discrete} optimization problem to optimize the \textit{weight coefficients} adopted at the source and relay to maximize the minimum Euclidean distance of the composite constellation of the received signal at the destination, subject to their individual average power constraints. By taking advantage of the \textit{punched Farey sequence} and its unique properties, we attained  the optimal solution to the formulated problem. Simulation results demonstrated the effectiveness of our proposed design based on SC over that using conventional time division multiple access (TDMA).


\begin{thebibliography}{00}
\bibitem{b1} J. N. Laneman, D. N. C. Tse and G. W. Wornell, ``Cooperative diversity in wireless networks: Efficient protocols and outage behavior," \textit{IEEE Trans. Inf. Theory}, vol. 50, no. 12, pp. 3062-3080, Dec. 2004.
\bibitem{b2} T. Cover and A. E. Gamal, ``Capacity theorems for the relay channel," \textit{IEEE Trans. Inf. Theory}, vol. 25, no. 5, pp. 572-584, Sept. 1979.
\bibitem{b3}  E. G. Larsson and B. R. Vojcic, ``Cooperative transmit diversity based on superposition modulation," \textit{IEEE Commun. Lett.}, vol. 9, no. 9, pp. 778-780, Sept. 2005.
\bibitem{b4} T. M. Cover, ``Comments on broadcast channels," \textit{IEEE Trans. Inf. Theory}, vol. 44, no. 6, pp. 2524-2530, Oct. 1998.
\bibitem{b5} A. Lozano, A. M. Tulino and S. Verdu, ``Optimum power allocation for parallel Gaussian channels with arbitrary input distributions," \textit{IEEE Trans. Inf. Theory}, vol. 52, no. 7, pp. 3033-3051, Jul. 2006.
\bibitem{b6} Y. Ding, J. Zhang and K. M. Wong, ``Optimal precoder for amplify-and-forward half-duplex relay system," \textit{IEEE Trans. Wireless Commun.}, vol. 7, no. 8, pp. 2890-2895, Aug. 2008.
\bibitem{b7} X. Jin and H. Kim, ``Switched-power two-layer superposition coding in cooperative decode-forward relay systems," \textit{IEEE Trans. Wireless Commun.}, vol. 15, no. 3, pp. 2193-2204, Mar. 2016.
\bibitem{b8} L. J. Rodríguez, N. H. Tran and T. Le-Ngoc, ``High-rate BICM-ID with superposition modulation over amplify-and-forward relay channels," in \textit{2010 25th Biennial Symposium on Commun.}, Kingston, ON, 2010, pp. 150-154.
\bibitem{b9} Z. Dong, H. Chen, J. Zhang, L. Huang and B. Vucetic, ``Uplink non-orthogonal multiple access with finite-alphabet inputs," \textit{IEEE Trans. Wireless Commun.}, vol. 17, no. 9, pp. 5743-5758, Sept. 2018.
\bibitem{b10} A. Goldsmith, \textit{Wireless Communications}. Cambridge University Press, 2005.
\bibitem{b11} V. R. Cadambe, S. A. Jafar and C. Wang, ``Interference alignment with asymmetric complex signaling—Settling the Høst-Madsen–Nosratinia conjecture," \textit{IEEE Trans. Inf. Theory}, vol. 56, no. 9, pp. 4552-4565, Sept. 2010.
\bibitem{b12} J. Harshan and B. S. Rajan, ``On two-user Gaussian multiple access channels with finite input constellations," \textit{IEEE Trans. Inf. Theory,} vol. 57, no. 3, pp. 1299-1327, Mar. 2011.
\bibitem{b13}  X. Xiao, Q. Huang and E. Viterbo, ``Joint optimization scheme and sum constellation distribution for multi-user Gaussian multiple access channels with finite input constellations," in \textit{2016 Austral. Commun. Theory Workshop (AusCTW)}, Melbourne, VIC, 2016, pp. 130-135.
\bibitem{b14} Z. Dong, H. Chen, J. Zhang and L. Huang, ``On non-orthogonal multiple access with finite-alphabet inputs in Z-channels," \textit{IEEE J. Sel. Areas Commun.}, vol. 35, no. 12, pp. 2829-2845, Dec. 2017.


\end{thebibliography}
\end{document}